\documentclass[11pt]{revtex4}    % Specifies the document style.
\usepackage{amssymb,epsf}
\usepackage{latexsym}
\usepackage{graphicx}
\begin{document}

\title{Wormhole Solutions in Gauss-Bonnet-Born-Infeld Gravity}
\author{M. H. Dehghani$^{1,2}$\footnote{mhd@shirazu.ac.ir} and S. H.
Hendi$^{1,3}$\footnote{hendi@mail.yu.ac.ir}}

\address{$^1$ Physics Department and Biruni Observatory, College of Sciences, Shiraz University, Shiraz 71454, Iran\\
         $^2$ Research Institute for Astrophysics and Astronomy of Maragha (RIAAM), Maragha, Iran\\
         $^3$ Physics Department, College of Sciences, Yasouj University, Yasouj 75914, Iran}

\begin{abstract}
A new class of solutions which yields
an $(n+1)$-dimensional spacetime with a longitudinal nonlinear magnetic
field is introduced. These spacetimes have no curvature singularity and no horizon,
and the magnetic field is non singular in the whole spacetime. They may be
interpreted as traversable wormholes which could be supported by matter not violating the weak energy
conditions. We generalize this class of
solutions to the case of rotating solutions and show that
the rotating wormhole solutions have a net electric charge which is
proportional to the magnitude of the rotation parameter, while the
static wormhole has no net electric charge. Finally, we use the
counterterm method and compute the conserved quantities of these
spacetimes.
\end{abstract}

\maketitle

\section{Introduction}

The intriguing possibility that our Universe is only a part of a higher
dimensional spacetime has raised a lot of interest in physics community
recently. In string theory extra dimensions are necessary since superstring
theory requires a ten-dimensional spacetime to be consistent from the
quantum point of view. The effect of string theory on the left hand side of
field equations of gravity is usually investigated by means of a low energy
effective action which describes gravity at the classical level \cite{Wit}.
This effective action consists of the Einstein-Hilbert action plus
curvature-squared terms and higher powers as well, and in general give rise
to fourth order field equations and bring in ghosts. However, if the
effective action contains the higher powers of curvature in particular
combinations, then only second order field equations are produced and
consequently no ghosts arise \cite{Zum}. The effective action obtained by
this argument is precisely of the form proposed by Lovelock \cite{Lov}. On
the other hand, the dynamics of D-branes and some soliton solutions of
supergravity is governed by the Born-Infeld action \cite{Gibbons}, and
therefore if one want to couple electromagnetic field with gravity it is
more suitable to put the energy momentum of Born-Infeld electromagnetic
field on the right hand side of the field equations. While the Lovelock
gravity was proposed to have field equations with at most second order
derivatives of the metric \cite{Lov}, the nonlinear electrodynamics
proposed, by Born and Infeld, with the aim of obtaining a finite value for
the self-energy of a point-like charge \cite{BI}. The Lovelock gravity
reduces to Einstein gravity in four dimensions and also in the weak field
limit, while the Lagrangian of the Born-Infeld (BI) electrodynamics reduces
to the Maxwell Lagrangian in the weak field limit. Considering the analogy
between the Lovelock and the Born-Infeld terms, which are on similar footing
with regard to string corrections on the gravity side and electrodynamic
side, respectively, it is plausible to include both these corrections
simultaneously. Here we restrict ourself to the first three terms of
Lovelock gravity. The first two terms are the Einstein-Hilbert term with
cosmological constant, while the third term is known as the Gauss-Bonnet
term. In recent decades, the exact solutions of Einstein-Gauss-Bonnet
gravity and their properties were studied. For example, static spherically
symmetric black hole solutions of the Gauss-Bonnet gravity were found in
Ref. \cite{Des}. Black hole solutions with nontrivial topology in this
theory were also studied in Refs. \cite{Cai,Aros,Ish}. NUT charged black
hole solutions of Gauss-Bonnet gravity and Gauss-Bonnet-Maxwell gravity were
obtained in \cite{DMH}. Also rotating black hole solutions of
Gauss-Bonnet-Maxwell and Born-Infeld-Gauss-Bonnet gravity have been studied
in \cite{Deh1,Hendi2}.

Currently, there exist some activities in the field of wormhole physics
following, particularly, the seminal works of Morris, Thorne and Yurtsever
\cite{MorTho}. Morris and Thorne assumed that their traversable wormholes
were time independent, non-rotating, and spherically symmetric bridges
between two universes. The manifold of interest is thus a static spherically
symmetric spacetime possessing two asymptotically flat regions. These kinds
of {\bf \ }wormholes could be threaded both by quantum and classical matter
fields that violate certain energy conditions at least at the throat known
as exotic matter. On general grounds, it has recently been shown that the
amount of exotic matter needed at the wormhole throat can be made
arbitrarily small thereby facilitating an easier construction of wormholes
\cite{VisDad}. Lorentzian wormholes in spacetimes with more than four
dimensions were analyzed by several authors \cite{KarSah}. In particular,
wormholes in Gauss-Bonnet gravity were considered in Ref. \cite{BhaKar}. In
this paper we are looking for the $(n+1)$-dimensional horizonless solutions
in Born-Infeld-Gauss-Bonnet gravity. The motivation for constructing these
kinds of solutions is that they may be interpreted as wormhole. Here, we
will find these kinds of horizonless solutions in the
Born-Infeld-Gauss-Bonnet gravity, and use the counterterm method to compute
the conserved quantities of the system. The outline of our paper is as
follows. In Sec. \ref{Long} we give a brief review of the field equations of
Born-Infeld-Gauss-Bonnet gravity, present a new class of static wormhole
solutions which produces longitudinal magnetic field, and then we consider
the properties of the solutions as well as the weak energy condition. In
Sec. \ref{Angul} we endow these spacetime with global rotations and apply
the counterterm method to compute the conserved quantities of these
solutions. Finally, we finish our paper with some closing remarks.

\section{Static Wormhole Solutions \label{Long}}

We begin this section with a brief review of Einstein-Gauss-Bonnet gravity
in the presence of nonlinear Born-Infeld electromagnetic field. The field
equations of this theory in the presence of a negative cosmological constant may be written as
\begin{equation}
G_{\mu \nu }-\frac{n(n-1)}{2l^2} g_{\mu \nu }+\alpha H_{\mu \nu }=\frac{1}{2}g_{\mu \nu
}L(F)+\frac{2F_{\mu \lambda }F_{{\nu }}\,^{\lambda }}{\sqrt{1+\frac{F^{2}}{%
2\beta ^{2}}}},  \label{Geq}
\end{equation}
\begin{equation}
\partial _{\mu }\left( \frac{\sqrt{-g}F^{\mu \nu }}{\sqrt{1+\frac{F^{2}}{%
2\beta ^{2}}}}\right) =0,  \label{Maxeq}
\end{equation}
where $G_{\mu \nu }$ is the Einstein tensor and $H_{\mu \nu }$\ is the
divergence-free symmetric tensor
\begin{eqnarray}
H_{\mu \nu }&=&2R_{\mu }^{\ \rho
\sigma \lambda }R_{\nu \rho \sigma \lambda }-4R^{\rho \sigma }R_{\mu \rho \nu \sigma }+2RR_{\mu \nu }-4R_{\mu \lambda
}R_{\text{ \ }\nu }^{\lambda }  \nonumber \\
&&-\frac{1}{2}g_{\mu \nu }\left(R_{\kappa \lambda \rho \sigma }R^{\kappa \lambda
\rho \sigma }-4R_{\rho \sigma }R^{\rho \sigma }+R^{2}\right).
\end{eqnarray}
In Eq. (\ref{Geq}) $\beta $\ is the Born-Infeld parameter with dimension of
mass, $F^{2}=F^{\mu \nu }F_{\mu \nu }$\ where $F_{\mu \nu }$\ is
electromagnetic field{\bf \ }tensor, and $L(F)$ is the Born-Infeld
Lagrangian given as
\begin{equation}
L(F)=4\beta ^{2}\left( 1-\sqrt{1+\frac{F^{2}}{2\beta ^{2}}}\right).
\end{equation}
In the limit $\beta \rightarrow \infty $, $L(F)$ reduces to the standard
Maxwell form $L(F)=-F^{2}$, while $L(F)\rightarrow 0$ as $\beta \rightarrow
0 $. Equation (\ref{Geq}) does not contain the derivative of the curvatures,
and therefore the derivatives of the metric higher than two do not appear.

Here, we want to obtain the $(n+1)$-dimensional solutions of Eqs. (\ref{Geq}%
) and (\ref{Maxeq}) which produce longitudinal magnetic fields in the
Euclidean submanifold spans by the $x^{i}$ coordinates ($i=1,...,n-3$). We
assume that the metric has the following form:
\begin{equation}
ds^{2}=-\frac{r^{2}}{l^{2}}dt^{2}+\frac{dr^{2}}{f(r)}+\Gamma
^{2}l^{2}f(r)d\psi ^{2}+r^{2}d\phi ^{2}+\frac{r^{2}}{l^{2}}dX^{2},
\label{Met1a}
\end{equation}
where $dX^{2}={{\sum_{i=1}^{n-3}}}(dx^{i})^{2}$ and $\Gamma $ is a constant
will be fixed later. Note that the coordinates $-\infty <x^{i}<\infty $ have
the dimension of length, while the angular coordinates $\psi $ and $\phi $
are dimensionless as usual and range in $[0,2\pi ]$. The motivation for this
metric gauge $[g_{tt}\varpropto -r^{2}$ and $(g_{rr})^{-1}\varpropto g_{\psi
\psi }]$ instead of the usual Schwarzschild gauge $[(g_{rr})^{-1}\varpropto
g_{tt}]$ comes from the fact that we are looking for a horizonless magnetic
solution. The electromagnetic field equation (\ref{Maxeq}) reduces to
\begin{equation}
\beta ^{2}l^{2}\Gamma ^{2}\left[ rF_{\psi r}^{^{\prime
}}(r)+(n-1)F_{\psi r}(r)\right] +(n-1)F_{\psi r}^{3}(r)=0,
\label{Fpr eq}
\end{equation}%
where the prime denotes a derivative with respect to the $r$
coordinate. The solution of Eq. (\ref{Fpr eq}) can be written as
\begin{equation}
F_{\psi r}=\frac{2\Gamma ^{2}l^{n-2}q}{r^{n-1}\sqrt{1-\eta }},  \label{Fpr}
\end{equation}
where $q$ is an arbitrary constant and
\begin{equation}
\eta ={\frac{4\Gamma ^{2}q^{2}l^{2(n-3)}}{\beta ^{2}r^{2(n-1)}}}.  \label{eta}
\end{equation}
Equation (\ref{Fpr}) shows that $r$ should be greater than $r_{01}=(2\Gamma
ql^{n-3}/\beta )^{1/(n-1)}$ in order to have a real nonlinear
electromagnetic field and consequently a real spacetime. To find the
function $f(r)$, one may use any components of Eq. (\ref{Geq}). The simplest
equation is the $rr$ component of these equations which can be written as
\begin{eqnarray}
&&(n-1)\{l^{2}r^{n-4}[2(n-2)(n-3)\alpha f-r^{2}]f^{^{\prime
}}+(n-2)l^{2}r^{n-5}[(n-3)(n-4)\alpha f-r^{2}]f  \nonumber \\
&&+nr^{n-1}\}+4\beta ^{2}l^{2}r^{n-1}(1-\sqrt{1-\eta })=0.  \label{rrcom}
\end{eqnarray}
The solution of Eq. (\ref{rrcom}), which also satisfies all the other components of
the gravitational field equations (\ref{Geq}), can be written as
\begin{equation}
f(r)={\frac{r^{2}}{2\,(n-2)\,(n-3)\,\alpha }}\left( 1-\sqrt{g(r)}\right),  \label{f(r)}
\end{equation}
where
\begin{eqnarray}
g(r) &=&1+16\frac{(n-3)\alpha \beta ^{2}\eta }{n}\text{ }%
_{2}F_{1}\left( \left[ \frac{1}{2}{,}\frac{n{-2}}{2n{-2}}\right] ,\left[
\frac{{3}n{-4}}{{2}n{-2}}\right] ,{\eta }\right) \nonumber\\
&&+ 4(n-2)(n-3)\alpha \left[ \frac{4\beta ^{2}\left(
\sqrt{1-\eta }-1\right)}{n(n-1)} -\frac{1}{l^{2}}-
\frac{2ml}{r^{n}} \right],  \label{gr}
\end{eqnarray}
In Eq. (\ref{gr}) $m$ is an integration constant which is related to geometrical mass of
the spacetime and $_{2}F_{1}([a,b],[c],z)$ is the hypergeometric function which may be defined as \cite{Handbook}
\[
_{2}F_{1}\left( %
\left[ \frac{1}{2}{,}\frac{n{-2}}{2n{-2}}\right] ,\left[ \frac{{3}n{-4}}{{2}n%
{-2}}\right] ,{b u^{2n-2} }\right)=\frac{n-2}{u^{n-2}}\int \frac{u^{n-3}}{\sqrt{1-b u^{2n-2}}}du.
\]
One may note that the above asymptotically AdS solution reduces to Einstein-Born-Infeld solution when
$\alpha $ vanishes and reduces to the solution introduced in \cite{Deh2} as $%
\beta $ goes to infinity. When $m$ and $q$ are zero, the vacuum solution is
\begin{equation}
f(r)=\frac{r^{2}}{2(n-2)(n-3)\alpha }\left( 1-\sqrt{1-\frac{%
4(n-2)(n-3)\alpha }{l^{2}}}\right).  \label{Fg0}
\end{equation}
Equation (\ref{Fg0}) shows that for a positive value of $\alpha $, this
parameter should satisfies $\alpha \leq l^{2}/4(n-2)(n-3)$. Also note that
$l_{{\rm eff}}$ for the AdS solution of the theory is
\begin{equation}
l_{{\rm eff}}=[2(n-2)(n-3)\alpha]^{1/2} \left( 1-\sqrt{1-\frac{4(n-2)(n-3)\alpha
}{l^{2}}}\right) ^{-1/2},  \label{leff}
\end{equation}
which reduces to $l$ as $\alpha $ goes to zero.

\begin{figure}[ht]
\centering
{\ \includegraphics[width=7cm]{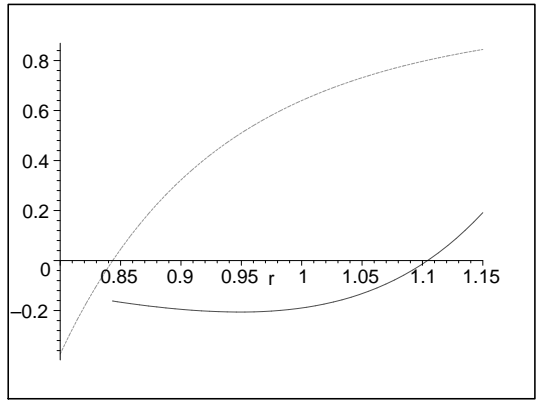} }
\caption{The functions $g(r)$ (continuous-line) and $1-\eta$ (dashed-line) versus $r$ for $n=4$, $q =0.3$, $m=0.8$, $l=\Gamma=\beta=1$, and $\alpha=0.2$
.}
\label{Fig1}
\end{figure}

To have a real spacetime, the function $g(r)$ should be positive. This occurs provided the mass parameter $m\leq m_0$, where $m_0$ is the value of $m$ calculated from the equation
$g(r=r_{01})=0$ given as:
\[
m_{0}=\frac{r_{01}^{n}}{2 l}\left\{ \frac{1}{4(n-2)(n-3)\alpha}-\frac{1}{l^{2}}
+\frac{4\beta ^{2}}{n(n-2)}\text{ }_{2}F_{1}\left( \left[ \frac{1}{2}{,}\frac{n{-2}}{%
2n{-2}}\right] ,\left[ \frac{{3}n{-4}}{{2}n{-2}}\right] ,{1}\right)
-\frac{4\beta ^{2}}{n(n-1)}\right\}
\]
In this case the metric function $f(r)$ is real for $r \geq r_0=r_{01}$. For the case of $m > m_{0}$,
$r$ should be greater than $r_{0}$ in order to have a real spacetime,
where $r_{0}$ is the largest real root of $g(r)=0$. Figure \ref{Fig1} shows
the zeros of $g(r)$ and $1-\eta$ ($r_{0}$ and $r_{01}$ respectively) for $m>m_0$.

In order to study the general structure of this solution, we first look for
curvature singularities. It is easy to show that the Kretschmann invariant $%
R_{\mu \nu \lambda \kappa }R^{\mu \nu \lambda \kappa }$\ diverges at $%
r=r_{0} $, it is finite for $r>r_{0}$\ and goes to zero as $r\rightarrow
\infty $. Therefore one might think that there is a curvature singularity
located at $r=r_{0}$. Two cases happen. In the first case the function $f(r)$%
has no real root greater than $r_{0}$, and therefore we encounter with a
naked singularity which we are not interested in it. So we consider only the
second case which the function has one or more real root(s) larger than $%
r_{0}$. In this case the function $f(r)$\ is negative for $r<r_{+}$, and
positive for $r>r_{+}$\ where $r_{+}$\ is the largest real root of $f(r)=0$.
Indeed, $g_{rr}$\ and $g_{\psi \psi }$\ are related by $f(r)=g_{rr}^{-1}=%
\Gamma ^{-2}l^{-2}g_{\psi \psi }$, and therefore when $g_{rr}$\ becomes
negative (which occurs for $r_{0}<r<r_{+}$) so does $g_{\psi \psi }$. This
leads to an apparent change of signature of the metric from $(n-1)+$\ to $%
(n-2)+$, and therefore indicates that $r$ should be greater than $r_{+}$.
Thus the\ coordinate $r$\ assumes the value $r_{+}\leq r<$\ $\infty $. The
function $f(r)$ given in Eq. (\ref{f(r)}) is positive in the whole spacetime
and is zero at $r=r_{+}$.
The Kretschmann scalar is a linear combination of the square of $f''$, $f'/r$ and $f/r^2$. Since these
terms do not diverge in the range $%
r_{+}\leq r<\infty $, one finds that the Kretschmann scalar is finite. Also one can show that
other curvature invariants (such as Ricci
scalar, Ricci square, Weyl square and so on) are functions of $f''$, $f'/r$ and $f/r^2$
and therefore the spacetime has no curvature singularity in the range $r_{+}\leq r<\infty $. In order to avoid conic singularity at $r=r_{+}$ in
the $(r-\psi )$-section, one may fix the factor $\Gamma =1/[lf^{\prime
}(r_{+})]$ in the metric.

Now, we investigate the wormhole interpretation of the above solution. It
satisfies the so-called flare-out condition when $r=r_{+}$. This can be seen
by embedding the 2-surface of constant $t$, $\psi $ and $x^{i}$'s with the
metric
\begin{equation}
ds^{2}=\frac{dr^{2}}{f(r)}+r^{2}d\phi ^{2},  \label{metrflare}
\end{equation}
in an (unphysical) three-dimensional Euclidean flat space, which has the
metric
\begin{equation}
ds^{2}=dr^{2}+r^{2}d\phi ^{2}+dz^{2}
\end{equation}
in cylindrical coordinates. The surface described by the function $z=z(r)$
satisfies
\begin{eqnarray}
\frac{dr}{dz} &=&\sqrt{\frac{f(r)}{1-f(r)}}=0,  \label{dzdr} \\
\frac{d^{2}r}{dz^{2}} &=&\frac{f^{\prime }}{2\left[ 1-f\right] ^{2}}>0,
\end{eqnarray}
which show that it has the characteristic shape of a wormhole, as
illustrated in Figs. 1 and 2 of Ref. \cite{MorTho}. Next we consider the
Weak Energy Condition (WEC) for the above solution. Using the orthonormal
contravariant (hatted) basis vectors
\[
{\bf e}_{\widehat{t}}=\frac{l}{r}\frac{\partial }{\partial t},\text{ \ \ }%
{\bf e}_{\widehat{r}}=f^{1/2}\frac{\partial }{\partial r},\text{ \ \ }{\bf e}%
_{\widehat{\psi }}=\frac{1}{\Gamma lf^{1/2}}\frac{\partial }{\partial \psi },%
\text{ \ }{\bf e}_{\widehat{\phi }}=r^{-1}\frac{\partial }{\partial \phi },%
\text{ \ }{\bf e}_{\widehat{x^{i}}}=\frac{l}{r}\frac{\partial }{\partial
x^{i}},
\]
the mathematics and physical interpretations become simplified. It is a
matter of straight forward calculations to show that the components of stress-energy
tensor are
\begin{eqnarray}
T_{_{\widehat{t}\widehat{t}}} &=&-T_{_{\widehat{\phi }\widehat{\phi }%
}}=-T_{_{\widehat{i}\widehat{i}}}=\frac{\beta ^{2}}{4\pi }\left( \left[
1+\left( \frac{F_{\psi r}}{\Gamma l\beta }\right) ^{2}\right]
^{1/2}-1\right) ,  \nonumber \\
\text{ \ \ }T_{_{\widehat{r}\widehat{r}}} &=&T_{_{\widehat{\psi }\widehat{%
\psi }}}=\frac{\beta ^{2}}{4\pi }\left( 1-\left[ 1+\left( \frac{F_{\psi r}}{%
\Gamma l\beta }\right) ^{2}\right] ^{-1/2}\right),  \label{EMtensor}
\end{eqnarray}
which satisfy the weak energy conditions:
\begin{equation}
T_{_{\widehat{t}\widehat{t}}}\geq 0,\hspace{1cm}T_{_{\widehat{t}\widehat{t}%
}}+T_{_{\widehat{i}\widehat{i}}}=T_{_{\widehat{t}\widehat{t}}}+T_{_{\widehat{%
\phi}\widehat{\phi}}}\geq 0,\text{ \ }T_{_{\widehat{t}\widehat{t}}}+T_{_{%
\widehat{r}\widehat{r}}}=T_{_{\widehat{t}\widehat{t}}}+T_{_{\widehat{\psi }%
\widehat{\psi }}}\geq 0.  \label{WEC}
\end{equation}

\section{Rotating Wormhole solutions\label{Angul}}

First, we want to endow our spacetime solution (\ref{Met1a}) with a global
rotation. In order to add angular momentum to the spacetime, we perform the
following rotation boost in the $t-\psi $ plane
\begin{equation}
t\mapsto \Xi t-a\psi ,\hspace{0.5cm}\psi \mapsto \Xi \psi -\frac{a}{l^{2}}t,
\label{Tr}
\end{equation}
where $a$ is a rotation parameter and $\Xi =\sqrt{1+a^{2}/l^{2}}$.
Substituting Eq. (\ref{Tr}) into Eq. (\ref{Met1a}) we obtain
\begin{equation}
ds^{2}=-\frac{r^{2}}{l^{2}}\left( \Xi dt-ad\psi \right) ^{2}+\frac{dr^{2}}{%
f(r)}+\Gamma ^{2}l^{2}f(r)\left( \frac{a}{l^{2}}dt-\Xi d\psi \right)
^{2}+r^{2}d\phi ^{2}+\frac{r^{2}}{l^{2}}dX^{2},  \label{Metr2}
\end{equation}
where $f(r)$ is the same as $f(r)$ given in Eq. (\ref{f(r)}). The
non-vanishing components of electromagnetic field tensor are now given by
\begin{equation}
F_{tr}=\frac{a}{\Xi l^{2}}F_{r\psi }=\frac{2\Gamma ^{2}l^{n-4}qa}{r^{n-1}%
\sqrt{1-\eta }}.
\end{equation}
Because of the periodic nature of $\psi$, the transformation (\ref{Tr}) is not a proper coordinate transformation
on the entire manifold. Therefore, the metrics (\ref{Met1a})
and (\ref{Metr2}) can be locally mapped into each other but not globally,
and so they are distinct \cite{Sta}. Note that this spacetime has no horizon and
curvature singularity. One should note that these solutions are different
from those discussed in \cite{Deh1}, which were electrically charged
rotating black brane solutions in Gauss-Bonnet gravity. The electric
solutions have black holes, while the magnetic solution interpret as
wormhole. It is worthwhile to mention that this solution reduces to the
solution of Einstein-Maxwell equation introduced in \cite{Deh2} as $\alpha $
goes to zero.

Second, we generalize the above solution to the case of rotating solution
with more rotation parameters. The rotation group in $n+1$ dimensions is $%
SO(n)$ and therefore the number of independent rotation parameters is $[n/2]$%
, where $[x]$ is the integer part of $x$. The generalized solution with $%
k\leq \lbrack n/2]$ rotation parameters can be written as
\begin{eqnarray}
ds^{2} &=&-\frac{r^{2}}{l^{2}}\left( \Xi dt-{{\sum_{i=1}^{k}}}a_{i}d\psi
^{i}\right) ^{2}+\Gamma ^{2}f(r)\left( \sqrt{\Xi ^{2}-1}dt-\frac{\Xi }{\sqrt{%
\Xi ^{2}-1}}{{\sum_{i=1}^{k}}}a_{i}d\psi ^{i}\right) ^{2}  \nonumber \\
&&+\frac{dr^{2}}{f(r)}+\frac{r^{2}}{l^{2}(\Xi ^{2}-1)}{\sum_{i<j}^{k}}%
(a_{i}d\psi _{j}-a_{j}d\psi _{i})^{2}+r^{2}d\phi ^{2}+\frac{r^{2}}{l^{2}}%
dX^{2},  \label{Metr5}
\end{eqnarray}
where $\Xi =\sqrt{1+\sum_{i}^{k}a_{i}^{2}/l^{2}}$, $dX^{2}$ is the Euclidean
metric on the $(n-k-2)$-dimensional submanifold and $f(r)$ is the same as $%
f(r)$ given in Eq. (\ref{f(r)}). The non-vanishing components of
electromagnetic field tensor are
\begin{equation}
F_{tr}=\frac{(\Xi ^{2}-1)}{\Xi a_{i}}F_{r\psi ^{i}}=\frac{2ql^{n-3}\Gamma
^{2}\sqrt{\Xi ^{2}-1}}{r^{n-1}\sqrt{1-\eta }}.
\end{equation}

\subsection{Conserved Quantities\label{Conserv}}

In general the conserved quantities are divergent when evaluated on the
solutions. A systematic method of dealing with this divergence for
asymptotically AdS solutions of Einstein gravity is through the use of the
counterterms method inspired by the anti-de Sitter conformal field theory
(AdS/CFT) correspondence \cite{Mal}. For asymptotically AdS solutions of
Lovelock gravity with flat boundary, $\widehat{R}_{abcd}(\gamma )=0$, the
finite energy momentum tensor is \cite{DBSH,DM1}
\begin{equation}
T^{ab}=\frac{1}{8\pi }\{(K^{ab}-K\gamma ^{ab})+2\alpha (3J^{ab}-J\gamma
^{ab})-\left( \frac{n-1}{L}\right) \gamma ^{ab}\},  \label{Stress}
\end{equation}
where $L$ is
\begin{equation}
L=\frac{3 \sqrt{\chi} \left( 1-\sqrt{1-\chi }\right) ^{3/2}}{\sqrt{8} \left[%
1- \left( 1+\frac{\chi }{2}\right) \sqrt{1-\chi }\right] }l,  \label{L}
\end{equation}
and $\chi =4(n-2)(n-3)\alpha /l^{2}$. In Eq. (\ref{Stress}), $K^{ab}$ is the
extrinsic curvature of the boundary, $K$ is its trace, $\gamma ^{ab}$ is the
induced metric of the boundary, and $J$ is trace of $J^{ab}$
\begin{equation}
J_{ab}=\frac{1}{3}%
(K_{cd}K^{cd}K_{ab}+2KK_{ac}K_{b}^{c}-2K_{ac}K^{cd}K_{db}-K^{2}K_{ab}).
\end{equation}
One may note that when $\alpha $ goes to zero, the finite stress-energy
tensor (\ref{Stress}) reduces to that of asymptotically AdS solutions of
Einstein gravity with flat boundary.

To compute the conserved charges of the spacetime, we choose a spacelike
surface ${\cal B}$ in $\partial {\cal M}$ with metric $\sigma _{ij}$, and
write the boundary metric in ADM (Arnowitt-Deser-Misner) form:
\begin{equation}
\gamma _{ab}dx^{a}dx^{a}=-N^{2}dt^{2}+\sigma _{ij}\left( d\phi
^{i}+V^{i}dt\right) \left( d\phi ^{j}+V^{j}dt\right) ,
\end{equation}
where the coordinates $\phi ^{i}$ are the angular variables parameterizing
the hypersurface of constant $r$ around the origin, and $N$ and $V^{i}$ are
the lapse and shift functions respectively. When there is a Killing vector
field ${\cal \xi }$ on the boundary, then the quasilocal conserved
quantities associated with the stress tensors of Eq. (\ref{Stress}) can be
written as
\begin{equation}
{\cal Q}({\cal \xi )}=\int_{{\cal B}}d^{n-1}\varphi \sqrt{\sigma }T_{ab}n^{a}%
{\cal \xi }^{b},  \label{charge}
\end{equation}
where $\sigma $ is the determinant of the metric $\sigma _{ij}$, and $n^{a}$
is the timelike unit normal vector to the boundary ${\cal B}${.} In the
context of counterterm method, the limit in which the boundary ${\cal B}$
becomes infinite (${\cal B}_{\infty }$) is taken, and the counterterm
prescription ensures that the action and conserved charges are finite. No
embedding of the surface ${\cal B}$ in to a reference of spacetime is
required and the quantities which are computed are intrinsic to the
spacetimes.

For our case, the magnetic solutions of Gauss-Bonnet gravity, the first
Killing vector is $\xi =\partial /\partial t$, therefore its associated
conserved charge is the total mass of the wormhole per unit volume $V_{n-k-2}
$, given by
\begin{equation}
M=\int_{{\cal B}}d^{n-1}x\sqrt{\sigma }T_{ab}n^{a}\xi ^{b}=\frac{(2\pi )^{k}%
}{4}\left[ n(\Xi ^{2}-1)+1\right] \Gamma m. \label{Mas}
\end{equation}
For the rotating solutions, the conserved quantities associated with the
rotational Killing symmetries generated by $\zeta _{i}=\partial /\partial
\phi ^{i}$ are the components of angular momentum per unit volume $V_{n-k-2}$
calculated as
\begin{equation}
J_{i}=\int_{{\cal B}}d^{n-1}x\sqrt{\sigma }T_{ab}n^{a}\zeta _{i}^{b}=\frac{%
(2\pi )^{k}}{4}\Gamma n\Xi ma_{i}.  \label{Ang}
\end{equation}
Next, we calculate the electric charge of the solutions. To determine the
electric field we should consider the projections of the electromagnetic
field tensor on special hypersurfaces. The normal to such hypersurfaces for
the spacetimes with a longitudinal magnetic field is
\[
u^{0}=\frac{1}{N},\hspace{0.5cm}u^{r}=0,\hspace{0.5cm}u^{i}=-\frac{N^{i}}{N},
\]
and the electric field is $E^{\mu }=g^{\mu \rho }F_{\rho \nu }u^{\nu }$.
Then the electric charge per unit volume $V_{n-k-2}$ can be found by
calculating the flux of the electromagnetic field at infinity, yielding
\begin{equation}
Q=\frac{(2\pi )^{k}}{2}\Gamma q\sqrt{\Xi ^{2}-1}.  \label{elecch}
\end{equation}
Note that the electric charge is proportional to the magnitude of rotation
parameters and is zero for the static solutions.

\section{CLOSING REMARKS}

Considering both the nonlinear invariant terms constructed by
electromagnetic field tensor ($F_{\mu \nu}F^{\mu \nu}$) and quadratic
invariant terms constructed by Riemann tensor ($R^2$, $R_{\mu \nu}R^{\mu
\nu} $ and $R_{\mu \nu \rho \sigma}R^{\mu \nu \rho \sigma}$) in action, we
have obtained a new class of rotating spacetimes in various dimensions, with
negative cosmological constant. These solutions are asymptotically anti-de
Sitter and reduce to the solutions of Gauss-Bonnet-Maxwell gravity when $%
\beta \rightarrow \infty $ and reduce to those of Einstein-Born-Infeld
gravity as $\alpha \rightarrow 0$. This class of solutions yields an $(n+1)$%
-dimensional spacetime with a longitudinal nonlinear and nonsingular
magnetic field (the only nonzero component of the vector potential is $%
A_{\phi}(r)$) generated by a static magnetic source. We have found that
these solutions have no curvature singularity and no horizons and may be
interpreted as a traversable wormhole near $r=r_{+}$. Also, we found that
the weak energy condition is not violated at the throat, which shows there
is no exotic matter near the throat.

Also we have generalized these solutions to the case of rotating spacetimes
with a longitudinal magnetic field. For the rotating wormhole, when a
rotational parameter is nonzero, the wormhole has a net electric charge
density which is proportional to the magnitude of the rotational parameter
given by $\sqrt{\Xi^{2}-1}$. For static case, the electric field vanishes,
and therefore the wormhole has no net electric charge. Finally, we applied
the counterterm method and calculated the conserved quantities of the
solutions. We have found that the conserved quantities do not depend on the
Born-Infeld parameter $\beta$.

\acknowledgments{This work has been supported by Research
Institute for Astrophysics and Astronomy of Maragha.}

\end{document}